%% file: main.tex
\documentclass[sigconf]{acmart}

\usepackage{cleveref}
\crefname{listing}{Configuration}{Configurations}
\usepackage{xspace}
\usepackage{tikz}
\usepackage{pgfplots}
\pgfplotsset{compat=1.18}
\usepackage{booktabs}
\usepackage{makecell}
\usepackage{pifont}
\usepackage{tabularx}
\usepackage{xcolor,colortbl}
\usepackage{multirow}
\usepackage{soul}
\usepackage{adjustbox}
\usepackage{tcolorbox}
\usepackage{longtable}
\usepackage{caption}
\usepackage{tablefootnote} 
\usepackage{float}
\floatstyle{plaintop}
\newfloat{listing}{thp}{lop} 
\floatname{listing}{Configuration} 

\usepackage{listings}

\lstdefinelanguage{yaml}{
  morekeywords={pipeline, name, operation, params, version, checksum, threshold, cores, test_ratio, val_ratio, seed, output_path, test, val, train},
  sensitive=false,
  morecomment=[l]{\#},
  morestring=[b]',
  morestring=[b]"
}

\newcommand{\cmark}{\ding{51}}%
\def\framework{\textsc{DataRec}\xspace}
\def\frameworkNoSpace{\textsc{DataRec}}

\AtBeginDocument{%
  }

\setcopyright{none}

\acmConference[SIGIR '25]{}{July 13--18,
  2025}{Padua, Italy}

\begin{document}

\title{\framework: A Python Library for Standardized and Reproducible Data Management in Recommender Systems}


\author{Alberto Carlo Maria Mancino}
\email{alberto.mancino@poliba.it}
\affiliation{%
  \institution{Politecnico di Bari}
  \city{Bari}
  \country{Italy}
}

\author{Salvatore Bufi}
\email{s.bufi@phd.poliba.it}
\affiliation{%
  \institution{Politecnico di Bari}
  \city{Bari}
  \country{Italy}
}

\author{Angela Di Fazio}
\email{angela.difazio@poliba.it}
\affiliation{%
  \institution{Politecnico di Bari}
  \city{Bari}
  \country{Italy}
}

\author{Antonio Ferrara}
\email{antonio.ferrara@poliba.it}
\affiliation{%
  \institution{Politecnico di Bari}
  \city{Bari}
  \country{Italy}
}

\author{Daniele Malitesta}
\email{daniele.malitesta@centralesupelec.fr}
\affiliation{%
  \institution{Université Paris-Saclay, CentraleSupélec, Inria}
  \city{Gif-sur-Yvette}
  \country{France}
}

\author{Claudio Pomo}
\email{claudio.pomo@poliba.it}
\affiliation{%
  \institution{Politecnico di Bari}
  \city{Bari}
  \country{Italy}
}

\author{Tommaso Di Noia}
\email{tommaso.dinoia@poliba.it}
\affiliation{%
  \institution{Politecnico di Bari}
  \city{Bari}
  \country{Italy}
}

\renewcommand{\shortauthors}{Mancino et al.}

\begin{abstract}
\input{sections/abstract}
\end{abstract}


\maketitle
\section*{Publication}
This paper was accepted for publication at ACM SIGIR 2025. This is the authors’ preprint version. Please cite the published version when available.
\input{sections/introduction}

\input{sections/sota}

\input{sections/frameworks}

\input{sections/xxx}

\input{sections/conclution}


\bibliographystyle{ACM-Reference-Format}
\bibliography{bibliography}

\appendix

\end{document}

%% file: sections/abstract.tex
Recommender systems have demonstrated significant impact across diverse domains, yet ensuring the reproducibility of experimental findings remains a persistent challenge. A primary obstacle lies in the fragmented and often opaque data management strategies employed during the preprocessing stage, where decisions about dataset selection, filtering, and splitting can substantially influence outcomes. To address these limitations, we introduce \framework, an open-source Python-based library specifically designed to unify and streamline data handling in recommender system research. By providing reproducible routines for dataset preparation, data versioning, and seamless integration with other frameworks, \framework promotes methodological standardization, interoperability, and comparability across different experimental setups. Our design is informed by an in-depth review of 55 state-of-the-art recommendation studies ensuring that \framework adopts best practices while addressing common pitfalls in data management. Ultimately, our contribution facilitates fair benchmarking, enhances reproducibility, and fosters greater trust in experimental results within the broader recommender systems community. The \framework library, documentation, and examples are freely available at \url{https://github.com/sisinflab/DataRec}.

%% file: sections/introduction.tex
\section{Introduction}


Over the past decade, recommender systems have become a pivotal area of research and innovation, driving transformative advancements across diverse domains, including e-commerce, entertainment, healthcare, and education. By tailoring content and services to individual preferences, these systems not only enhance user experiences but also generate significant economic value, establishing their development and evaluation as critical priorities for researchers and practitioners in both academia and industry.

Despite this progress, achieving fair, reliable, and reproducible results remains a persistent challenge in recommender system research. Addressing this challenge and avoiding biased comparisons, unreliable findings, and virtual scientific progress requires rigorous methodologies that encompass every aspect of the recommendation pipeline, from data management and model training to performance evaluation~\cite{DBLP:conf/recsys/DacremaCJ19, DBLP:conf/recsys/SunY00Q0G20, DBLP:journals/umuai/BelloginS21, DBLP:journals/csur/ZangerleB23}. 
Central to these issues is the principle of reproducibility, a cornerstone of scientific inquiry that enables the validation of findings, ensures fair comparisons, and fosters trust within the research community. To better understand the origins of these reproducibility concerns in recommender systems, it is crucial to examine the initial stages of the recommendation pipeline. Indeed, the data management phase emerges as a particularly influential factor in shaping the reliability and comparability of experimental results~\cite{DBLP:conf/recsys/SaidB14}.

The role of this phase extends far beyond simple preprocessing.
Instead, it encompasses a complex sequence of tasks, from dataset preparation and filtering to splitting and formatting, each of which plays a pivotal role in shaping the outcomes of experimental evaluations. 
While existing recommender system frameworks have undeniably contributed to improving the transparency and comparability of model training and evaluation~\cite{DBLP:journals/is/LopsPMSS23, DBLP:journals/jmlr/SalahTL20, DBLP:conf/recsys/SunY00Q0G20, DBLP:conf/cikm/ZhuLYZH21, DBLP:conf/aaai/KadiogluK24, DBLP:conf/cikm/ZhaoMHLCPLLWTMF21, DBLP:conf/cikm/Ekstrand20, DBLP:conf/sigir/WangZMLM20, DBLP:conf/www/ChiaTBHK22, DBLP:conf/recsys/MichielsVG22, DBLP:conf/recsys/GrahamMW19, DBLP:conf/sigir/AnelliBFMMPDN21}, their framework-specific implementations of data handling strategies become a significant source of concern. Although effective within their own ecosystems, these implementations often diverge significantly, leading to inconsistent experimental results and, critically, undermining the very foundation of comparability in research findings~\cite{DBLP:journals/sigir/Lin18, DBLP:journals/tois/DacremaBCJ21}. Adding to this challenge, the prevalence of unique, non-interchangeable data management strategies across different frameworks further exacerbates the problem, severely limiting cross-framework interoperability and, consequently, the potential for robust reproducibility. This pervasive fragmentation unequivocally highlights the urgent need for a unifying approach to data management, one that transcends the limitations of individual frameworks and fosters essential methodological consistency across the entire field.


To address all the above-mentioned challenges, we propose \textbf{\framework}, an open-source Python-based library specifically designed to align, unify, and streamline the data management phase of recommender system research.
Unlike traditional frameworks, which often aim to provide comprehensive solutions for the entire recommendation pipeline, \framework is a library designed to complement them by serving as a shared foundation for reproducible and interoperable data handling.
By unifying the diverse strategies currently employed across frameworks, \framework provides researchers with a comprehensive and open toolkit that facilitates both the standalone preprocessing of data and the coherent implementation of the pipelines of existing and new frameworks.
\frameworkNoSpace’s key features include support for widely used data formats, direct access to 18 built-in research datasets with explicit references, and reproducible pre-processing and splitting strategies. It can be easily integrated into standalone projects and offers interfaces for exporting datasets in formats compatible with other frameworks. By prioritizing interoperability and modularity, \framework streamlines best practices in data management, addresses a key challenge in the reproducibility of recommender systems, and simplifies the development process.

\framework design is guided by a systematic analysis of 55 recommendation papers published in leading venues between 2020 and 2024. This analysis identifies common strategies, inconsistencies, and challenges in data handling, and its findings directly inform the design of \framework, ensuring that the tool provides robust solutions to these issues while facilitating transparency, comparability, and replicability.
By addressing the often-overlooked yet critical data management phase, \framework advances the reproducibility of recommender system research, complementing existing frameworks and fostering more rigorous and trustworthy scientific practices. In particular, our contributions are threefold:  
\begin{itemize}
    \item A comprehensive survey of data-related practices in state-of-the-art recommender systems research, addressing prevalent challenges in dataset preparation, filtering, and splitting.
    \item A fine-grained analysis of how the most widely used frameworks support common data-related operations.
    \item The release of \framework, an open-source Python-based library for reproducible and interoperable data management in recommender systems research, publicly available at \url{https://github.com/sisinflab/DataRec}.  
\end{itemize}

%% file: sections/sota.tex
\section{Overview of Data Handling and Processing in Recommender Systems}

\input{tables/literature}
Managing and processing recommendation datasets is fundamental to recommender system research, with strategies that can vary from study to study. 
On the one hand, to identify the most commonly adopted approaches in state-of-the-art research, we surveyed recent papers published at top-tier conferences in search and recommendation and collected data-handling strategies involved in their proposed recommendation models. 
Then, on the other side, we critically analysed leading frameworks collecting several recommender system techniques and whether they integrate (or do not) these functionalities. These analyses offer an overview of current practices in data handling for recommendation and serve as the motivational basis for the functionalities implemented in \framework. We deepen into these two investigations in the following paragraphs.


\subsection{Current Trends in Recommender Research}
Techniques and procedures for data handling in recommender systems are diverse and may depend on the specific dataset, domain, study, or implementation. To identify the most commonly adopted approaches, we analyzed 55 papers published between 2020 and 2024, extracted from five surveys covering diverse domains in recommendation: \textbf{graph neural networks}~\cite{DBLP:journals/csur/WuSZXC23}, \textbf{contrastive learning}~\cite{DBLP:journals/tois/JingZZW24}, \textbf{temporal modeling}~\cite{DBLP:journals/umuai/BoginaKJBKT23}, \textbf{reinforcement learning}~\cite{DBLP:journals/csur/AfsarCF23}, and \textbf{generative approaches}~\cite{DBLP:journals/corr/abs-2404-00579}. These papers were selected from the cited surveys to ensure they are relevant to the community, come from diverse domains, and represent state-of-the-art research. 
\Cref{tab:papers_datasets_complete} summarizes the results of our survey.

Acknowledging the diversity of existing approaches, our paper provides a comprehensive overview of current practices and their shortcomings, setting the stage for a unified approach to data management. 
In subsequent sections, we detail our findings in each key area, dataset usage, filtering strategies, and data-splitting methods and discuss their broader implications for the reproducibility of recommender systems research. This systematic investigation not only underscores the pressing need for methodological consistency but also motivates the development of \framework as a solution to these challenges.
Below, we present the results of our study, highlighting the differences in dataset selection, binarization and filtering techniques adopted, and splitting strategies for model evaluation.

\subsubsection{Datasets}
\input{tables/datasets_inpaper}
The surveyed papers reveal a remarkable breadth in dataset usage, with \textbf{79 distinct datasets} identified. 
Within these, the ones appearing in at least three different datasets are explicitly reported in~\Cref{tab:papers_datasets_complete}.
This extensive utilization of diverse datasets underscores the inherent adaptability and domain-agnostic nature of recommender systems, showcasing their capacity to address a wide array of real-world applications. 
Indeed, this heterogeneity reflects the field's maturity and its responsiveness to varied application contexts. 

However, our review raises a significant issue in how recommendation datasets are referenced, as evidenced by the further analysis reported in  \Cref{tab:dataset_references}.
While papers using open datasets should reference the \textbf{original data source}, only 35.2\% of the 179 dataset usages analyzed provide such an explicit reference.
Among the remaining cases, 32.4\% fail to cite the original source entirely, instead offering a \textbf{copy of a new version the dataset}---often already processed---without clarifying its origin. Although this practice may support reproducibility, it does not ensure a clear reproduction of the filtering methods applied, compromising the validation of these approaches.
Additionally, 16.8\% cite only the \textbf{dataset's original paper} or \textbf{other scholarly papers}, which does not guarantee clarity about the dataset used, especially when multiple online versions exist.
Finally, 15.6\% of the datasets are irretrievable, either due to the \textbf{absence of any reference} or because the provided \textbf{links are broken}, e.g, they are no longer functional or have been blocked.

This analysis highlights the importance of tracking and versioning recommendation datasets. By maintaining a direct reference to the original or trusted sources and by providing tools to reproduce filtering and splitting approaches, complete reproducibility can be ensured. For this reason, \framework includes classes and methods for direct and traceable management of the most widely used datasets in the analyzed papers, specifically, those appearing in at least three different articles, as reported in~\Cref{tab:papers_datasets_complete}.

\subsubsection{Binarization and Filtering}
Pre-processing strategies are critical in every experimental setting, as meaningful evaluations require aligning datasets with the experimental design. Consequently, ensuring the reproducibility of these algorithms is essential for valid comparisons across different experiments. Our analysis identified the most commonly adopted strategies --- retaining only those that appear in at least three datasets --- which are reported in~\Cref{tab:papers_datasets_complete}.

One fundamental pre-processing strategy is \textbf{binarization}, which converts explicit feedback into implicit feedback. 
This transformation is designed to facilitate the training and evaluation of models that predict the likelihood of interactions rather than precise rating values. This task has become increasingly prevalent in contemporary recommender systems~\cite{DBLP:conf/uai/RendleFGS09,DBLP:conf/icdm/HuKV08,DBLP:conf/recsys/TakacsPT11}. 
The \textbf{$k$-Core} method is widely employed to remove cold users (\textbf{$k$-Core User}), items (\textbf{$k$-Core Item}), or iteratively both of them (\textbf{$k$-Core Iterative}). Indeed, due to the lack of detailed preference information, these interactions are often considered noisy and are removed to prevent them from negatively affecting model performance.
Other strategies we identified are primarily associated with session-based recommendation experiments, where it could be useful to filter data based on the \textbf{session length}, in order to retain neither short nor long sessions.
Due to their wide adoption, we have included these techniques in \framework and other classical approaches. 
However, we do not include session-based approaches because they are not yet supported within the library.

\subsubsection{Splitting}
Data splitting strategies partition datasets for training, evaluating, and testing models.
Our survey confirms prior findings~\cite{DBLP:conf/recsys/MengMMO20, DBLP:conf/sigir/Sun23} that there is no universally accepted splitting method. 
Instead, studies employ diverse strategies, 
contributing to inconsistent rankings of recommender systems, even when the same datasets and metrics are used~\cite{DBLP:conf/recsys/DacremaCJ19, DBLP:journals/csur/ZhangYST19}.

The most commonly strategies, as reported in~\Cref{tab:papers_datasets_complete} include:
\begin{itemize}
    \item \textbf{Random Holdout Splitting}: Randomly partitions the dataset into training, validation, and test sets based on predefined percentages. This is the most common approach.
    \item \textbf{Temporal Leave-1-Out}: It is the second most common. Extracts the last transaction per user for test and one for validation.
    \item \textbf{Random Leave-1-Out}: Similarly, the temporal version extracts one transaction per user randomly chosen from the user history.
    \item \textbf{Pre-computed}: One common approach consists of releasing a copy of the train, test and validation splits used for the experiments.
    \item \textbf{Temporal Fixed}: Splits data based on timestamps, either globally (e.g., interactions after a fixed time point) or per user (e.g., a percentage of the most recent interactions).
\end{itemize}
Like the aforementioned filtering strategies, these have also been included in \framework along with other classic approaches.



%% file: tables/literature.tex
\begin{table*}[]
\centering
\caption{Datasets, pre-processing and splitting strategies adopted in 55 research papers published between 2020 and 2024. Only datasets and strategies appearing in at least 3 research papers are shown.}
\label{tab:papers_datasets_complete}
  \rowcolors{2}{white}{gray!25}
  \setlength{\tabcolsep}{3.3pt}
  \footnotesize
\begin{tabular}{*{3}{l}@{\hskip 1.5em}*{16}{c}@{\hskip 1.5em}*{7}{c}@{\hskip 1.5em}*{7}{c}}
    \toprule
                                    &               &                                                     & \multicolumn{16}{c}{\textbf{Datasets}}                                                                                                                                                                                                                                                                                                                                                                                                                                                                                                                                                                                                                                                                                                                                                          & \multicolumn{7}{c}{\textbf{Pre-processing}}                                                                                                                                                                                                                                                                                                  & \multicolumn{7}{c}{\textbf{Splitting}}                                                                                                                                                                                                                                                                                                                                           \\
        \cmidrule(r){4-19} \cmidrule(lr){20-26} \cmidrule(l){27-33}                                    
                \textbf{Paper}   & \textbf{Venue} & \textbf{Year}                                      & \rotatebox{90}{\textbf{Yelp}} & \rotatebox{90}{\textbf{Amazon Books}} & \rotatebox{90}{\textbf{MovieLens 1M}} & \rotatebox{90}{\textbf{Amazon Beauty}} & \rotatebox{90}{\textbf{Last.fm}} & \rotatebox{90}{\textbf{Gowalla}} & \rotatebox{90}{\textbf{Amazon Toys and Games}} & \rotatebox{90}{\textbf{Tmall}} & \rotatebox{90}{\textbf{Amazon Games}} & \rotatebox{90}{\textbf{Yelp2018}} & \rotatebox{90}{\textbf{Alibaba-iFashion}} & \rotatebox{90}{\textbf{Amazon Clothing}} & \rotatebox{90}{\textbf{Amazon Sports and Outdoors}} & \rotatebox{90}{\textbf{Epinions}} & \rotatebox{90}{\textbf{MovieLens 20M}} & \rotatebox{90}{\textbf{\textit{Others}}} & \rotatebox{90}{\textbf{\textit{None}}} & \rotatebox{90}{\textbf{Binarization}} & \rotatebox{90}{\textbf{$k$-Core Iterative}} & \rotatebox{90}{\textbf{$k$-Core User}} & \rotatebox{90}{\textbf{$k$-Core Item}} & \rotatebox{90}{\textbf{Session length}} & \rotatebox{90}{\textbf{\textit{Others}}} & \rotatebox{90}{\textbf{Random Hold-Out}} & \rotatebox{90}{\textbf{Temporal Leave-1-out Hold-Out}} & \rotatebox{90}{\textbf{Fixed}} & \rotatebox{90}{\textbf{Random Leave-1-out Hold-Out}} & \rotatebox{90}{\textbf{Temporal Fixed Handcrafted}} & \rotatebox{90}{\textbf{\textit{Others}}} & \rotatebox{90}{\textbf{\textit{NA}}} \\
                   \midrule 
\citet{DBLP:journals/tois/HaoYZLC23}       & TOIS      & 2023 &        &        & \cmark &        &        & \cmark &        &        &        &        &        &        &        &        &        & \cmark & \cmark &        &        &        &        &        &        & \cmark &        &        &        &        &        &        \\
\citet{DBLP:conf/sigir/ShuaiZWSHWL22}      & SIGIR     & 2022 & \cmark &        &        &        &        &        & \cmark &        &        &        &        & \cmark &        &        &        & \cmark &        &        &        & \cmark &        &        &        & \cmark &        &        &        &        &        &        \\
\citet{DBLP:conf/kdd/Jiang0H23}            & KDD       & 2023 & \cmark &        &        &        & \cmark &        &        &        &        &        &        &        &        &        &        & \cmark & \cmark &        & \cmark &        &        &        &        & \cmark &        &        &        &        &        &        \\
\citet{DBLP:conf/sigir/YuY00CN22}          & SIGIR     & 2022 &        & \cmark &        &        &        &        &        &        &        & \cmark &        &        &        &        &        & \cmark & \cmark & \cmark &        &        &        &        &        & \cmark &        &        &        &        &        &        \\
\citet{DBLP:conf/www/XiaHHLYK23}           & WWW       & 2023 & \cmark & \cmark &        &        &        & \cmark &        &        &        &        &        &        &        &        &        &        & \cmark &        &        &        &        &        &        & \cmark &        &        &        &        &        &        \\
\citet{DBLP:conf/recsys/Hansen0MMBTL20}    & RecSys    & 2020 &        &        &        &        &        &        &        &        &        &        &        &        &        &        &        & \cmark & \cmark &        &        &        &        &        &        &        &        &        &        & \cmark &        &        \\
\citet{DBLP:conf/cikm/0158ZLYY22}          & CIKM      & 2022 &        &        &        & \cmark &        &        & \cmark &        &        &        &        &        &        &        &        & \cmark &        & \cmark & \cmark &        &        &        &        &        & \cmark &        &        &        &        &        \\
\citet{DBLP:journals/tcyb/FuAIZHQ22}       & Cyb IEEE  & 2021 &        &        & \cmark &        &        &        &        &        &        &        &        &        &        &        & \cmark & \cmark &        & \cmark &        & \cmark &        &        &        &        &        &        &        &        & \cmark &        \\
\citet{DBLP:conf/mm/Wu0CLHS023}            & MM        & 2023 & \cmark &        & \cmark & \cmark &        &        &        &        &        &        &        &        &        &        &        & \cmark &        & \cmark &        & \cmark &        &        &        &        & \cmark &        &        &        &        &        \\
\citet{DBLP:conf/sigir/WangXFL0C23}        & SIGIR     & 2023 & \cmark & \cmark & \cmark &        &        &        &        &        &        &        &        &        &        &        &        &        &        & \cmark &        &        &        &        &        & \cmark &        &        &        &        &        &        \\
\citet{DBLP:conf/sigir/RenXZY023}          & SIGIR     & 2023 &        & \cmark &        &        &        & \cmark &        & \cmark &        &        &        &        &        &        &        &        & \cmark &        &        &        &        &        &        &        &        & \cmark &        &        &        &        \\
\citet{DBLP:conf/cikm/ZhangGYGLY21}        & CIKM      & 2021 &        &        &        &        &        &        &        &        &        &        &        &        &        &        &        & \cmark & \cmark & \cmark &        &        &        &        & \cmark & \cmark &        &        &        &        &        &        \\
\citet{DBLP:conf/kdd/GuoSTGZLTH21}         & SIGKDD    & 2021 &        &        &        &        &        &        &        & \cmark &        &        &        &        &        &        &        & \cmark & \cmark &        &        &        &        &        &        &        &        &        & \cmark &        &        &        \\
\citet{DBLP:journals/tkde/ZhangWYLW23}     & TKDE      & 2022 &        &        &        & \cmark &        &        &        &        & \cmark &        &        &        &        &        &        & \cmark &        &        & \cmark &        &        &        &        &        & \cmark &        &        &        &        &        \\
\citet{DBLP:conf/sigir/DuYZZ00LS23}        & SIGIR     & 2023 &        &        & \cmark & \cmark &        &        & \cmark &        &        &        &        &        &        &        &        &        &        & \cmark &        &        &        &        &        &        & \cmark &        &        &        &        &        \\
\citet{DBLP:conf/icde/Xie0YYGO021}         & ICDE      & 2021 &        &        & \cmark & \cmark &        &        &        &        & \cmark &        &        &        &        &        & \cmark &        &        & \cmark &        &        &        &        & \cmark &        & \cmark &        &        &        &        &        \\
\citet{DBLP:journals/tkde/HaoZZLSXLZ23}    & TKDE      & 2023 &        &        &        &        &        &        & \cmark & \cmark &        &        &        &        &        &        &        &        &        &        &        & \cmark & \cmark &        &        &        & \cmark &        &        &        &        &        \\
\citet{DBLP:conf/recsys/0008P020}          & RecSys    & 2020 &        &        &        & \cmark &        &        &        & \cmark & \cmark &        &        &        &        &        &        & \cmark &        & \cmark & \cmark &        &        &        &        &        & \cmark &        &        &        &        &        \\
\citet{DBLP:conf/pkdd/JingZZYT22}          & PKDD      & 2022 &        &        &        &        &        &        &        &        &        &        &        &        &        &        &        & \cmark & \cmark &        &        &        &        &        &        & \cmark &        &        &        &        &        &        \\
\citet{DBLP:journals/fcsc/WuHWWCLX22}      & FCS       & 2022 &        & \cmark &        &        &        &        &        &        &        &        &        &        &        &        &        & \cmark &        &        &        & \cmark &        &        &        &        & \cmark &        &        &        &        &        \\
\citet{DBLP:journals/tois/ZhangSYXOZ20}    & TOIS      & 2020 &        &        &        & \cmark & \cmark &        &        &        & \cmark &        &        &        &        &        & \cmark & \cmark &        & \cmark & \cmark & \cmark & \cmark &        & \cmark &        & \cmark &        &        &        & \cmark &        \\
\citet{DBLP:conf/www/XiaHSX23}             & WWW       & 2023 & \cmark & \cmark &        &        &        & \cmark &        &        &        &        &        &        &        &        &        &        & \cmark &        &        &        &        &        &        & \cmark &        &        &        &        &        &        \\
\citet{DBLP:conf/wsdm/ChenHXWXL23}         & WSDM      & 2023 & \cmark &        &        &        &        &        &        &        &        &        &        &        &        & \cmark &        & \cmark & \cmark &        &        &        &        &        &        &        &        &        & \cmark &        &        &        \\
\citet{DBLP:conf/sigir/XiaHXZYH22}         & SIGIR     & 2022 & \cmark & \cmark &        &        &        &        &        &        &        &        &        &        &        &        &        & \cmark & \cmark &        & \cmark &        &        &        &        & \cmark &        &        &        &        &        &        \\
\citet{DBLP:journals/eswa/BoginaVKD22}     & ESwA      & 2022 &        &        &        &        &        &        &        &        &        &        &        &        &        &        &        & \cmark &        &        &        &        &        & \cmark & \cmark & \cmark &        &        &        &        &        &        \\
\citet{DBLP:conf/recsys/MancinoFBMNS23}    & RecSys    & 2023 &        &        & \cmark &        &        &        &        &        &        &        &        &        &        &        &        & \cmark &        & \cmark & \cmark &        &        &        &        & \cmark &        &        &        &        &        &        \\
\citet{DBLP:conf/sigir/YangHXL22}          & SIGIR     & 2022 &        & \cmark &        &        &        &        &        &        &        & \cmark &        &        &        &        &        & \cmark & \cmark &        &        &        &        &        &        &        &        & \cmark &        &        &        &        \\
\citet{DBLP:conf/kdd/YangHXH23}            & KDD       & 2023 &        &        &        &        & \cmark &        &        &        &        &        & \cmark &        &        &        &        & \cmark &        &        & \cmark &        &        &        &        & \cmark &        &        &        &        &        &        \\
\citet{DBLP:conf/wsdm/WangX0SLG023}        & WSDM      & 2023 &        & \cmark &        &        & \cmark &        &        &        &        &        &        &        &        &        &        & \cmark &        &        & \cmark &        &        &        &        & \cmark &        &        &        &        &        &        \\
\citet{DBLP:conf/aaai/HuangXXDXLBXLY21}    & AAAI      & 2021 & \cmark &        &        &        &        &        &        &        &        &        &        &        &        & \cmark &        & \cmark & \cmark &        &        &        &        &        &        &        &        &        & \cmark &        &        &        \\
\citet{DBLP:conf/sigir/TianXLYZ22}         & SIGIR     & 2022 & \cmark & \cmark & \cmark &        &        &        &        &        &        &        &        &        &        &        &        &        &        & \cmark & \cmark &        &        &        &        & \cmark &        &        &        &        &        &        \\
\citet{DBLP:conf/iclr/Cai0XR23}            & ICLR      & 2023 & \cmark & \cmark &        &        &        & \cmark &        & \cmark &        &        &        &        &        &        &        & \cmark & \cmark &        &        &        &        &        &        &        &        & \cmark &        &        &        &        \\
\citet{DBLP:conf/sigir/0001DWLZ020}        & SIGIR     & 2020 &        & \cmark &        &        &        & \cmark &        &        &        & \cmark &        &        &        &        &        &        & \cmark &        &        &        &        &        &        &        &        & \cmark &        &        &        &        \\
\citet{DBLP:conf/cikm/WangL0HMHFC22}       & CIKM      & 2022 &        & \cmark &        &        &        & \cmark & \cmark &        &        &        &        & \cmark &        &        &        &        &        &        &        &        & \cmark &        &        & \cmark &        &        &        &        &        &        \\
\citet{DBLP:conf/sigir/Zou0MWQ0C22}        & SIGIR     & 2022 &        &        & \cmark &        & \cmark &        &        &        &        &        &        &        &        &        &        & \cmark &        & \cmark &        &        &        &        &        & \cmark &        &        &        &        &        &        \\
\citet{DBLP:conf/www/WeiHXZ23}             & WWW       & 2023 &        &        &        &        &        &        &        &        &        &        &        &        & \cmark &        &        & \cmark & \cmark &        &        &        &        &        &        &        &        & \cmark &        &        &        &        \\
\citet{DBLP:conf/dasfaa/WuXZACZZLH22}      & DASFAA    & 2022 &        &        &        &        &        &        &        & \cmark &        &        &        &        &        &        &        &        &        &        & \cmark &        &        &        &        &        & \cmark &        &        &        &        &        \\
\citet{DBLP:journals/tois/LiZC23}          & SIGIR     & 2023 & \cmark &        &        &        &        &        &        &        &        &        &        &        &        &        &        & \cmark & \cmark &        &        &        &        &        &        &        &        &        &        &        & \cmark &        \\
\citet{DBLP:conf/nips/RajputMSKVHHT0S23}   & NIPS      & 2024 &        &        &        & \cmark &        &        & \cmark &        &        &        &        &        & \cmark &        &        &        &        &        &        & \cmark &        &        &        &        & \cmark &        &        &        &        &        \\
\citet{DBLP:conf/aaai/ChenWHZW20}          & AAAI      & 2020 &        & \cmark &        &        &        & \cmark &        &        &        &        &        &        &        &        &        &        &        &        & \cmark &        &        &        &        & \cmark &        &        &        &        &        &        \\
\citet{DBLP:conf/sigir/HadaMS21}           & SIGIR     & 2021 &        &        &        &        &        &        &        &        &        &        &        & \cmark &        &        &        & \cmark &        &        & \cmark &        &        &        &        &        &        &        &        &        & \cmark &        \\
\citet{DBLP:conf/cikm/ZhouWZZWZWW20}       & CIKM      & 2020 & \cmark &        &        & \cmark & \cmark &        & \cmark &        &        &        &        &        & \cmark &        &        & \cmark & \cmark &        &        &        &        &        &        &        & \cmark &        &        &        &        &        \\
\citet{DBLP:conf/cikm/0013YYSC21}          & CIKM      & 2022 &        &        &        &        &        &        &        & \cmark &        &        &        &        &        &        &        & \cmark &        &        &        &        & \cmark & \cmark &        &        &        &        &        & \cmark &        &        \\
\citet{DBLP:conf/sigir/WuWF0CLX21}         & SIGIR     & 2021 &        &        &        &        &        &        &        &        &        &        & \cmark &        &        &        &        &        &        &        &        &        &        &        & \cmark & \cmark &        &        &        &        &        &        \\
\citet{DBLP:conf/ijcai/GuWSX22}            & IJCAI     & 2022 & \cmark &        &        &        &        &        &        &        &        &        &        &        &        &        &        & \cmark & \cmark &        &        &        &        &        &        &        &        &        & \cmark &        &        &        \\
\citet{DBLP:conf/www/YuYLWH021}            & WWW       & 2021 & \cmark &        &        &        & \cmark &        &        &        &        &        &        &        &        &        &        & \cmark & \cmark & \cmark &        &        &        &        &        &        &        &        &        &        & \cmark &        \\
\citet{DBLP:conf/sigir/ChangGZHNSJ021}     & SIGIR     & 2021 &        &        &        &        &        &        &        &        &        &        &        &        &        &        &        & \cmark &        &        & \cmark & \cmark &        &        &        &        &        &        &        & \cmark &        &        \\
\citet{DBLP:journals/isci/LatifiMJ21}      & Inf. Sci. & 2021 &        &        &        &        & \cmark &        &        &        &        &        &        &        &        &        &        & \cmark &        &        &        & \cmark &        & \cmark & \cmark &        &        &        &        &        & \cmark &        \\
\citet{DBLP:journals/umuai/SymeonidisKZ20} & UMUAI     & 2020 &        &        &        &        &        &        &        &        &        &        &        &        &        &        &        & \cmark &        &        &        &        &        & \cmark &        &        &        &        &        &        & \cmark &        \\
\citet{DBLP:conf/sigir/SongWJZHH21}        & SIGIR     & 2021 & \cmark &        &        &        &        &        &        &        &        &        &        &        &        &        &        &        & \cmark &        &        &        &        &        &        & \cmark &        &        &        &        &        &        \\
\citet{DBLP:conf/kdd/YuY000H21}            & KDD       & 2022 & \cmark &        &        &        & \cmark &        &        &        &        &        &        &        &        &        &        & \cmark & \cmark & \cmark &        &        &        &        &        &        &        &        &        &        & \cmark &        \\
\citet{DBLP:conf/bigdataconf/BaiZWN20}     & ICBD      & 2020 &        &        &        &        &        &        &        &        &        &        &        &        &        & \cmark &        & \cmark & \cmark &        &        &        &        &        &        &        &        &        &        &        &        & \cmark \\
\citet{DBLP:conf/wsdm/GeLGXLZP0GOZ21}      & WSDM      & 2021 &        &        & \cmark &        &        &        &        &        &        &        &        &        &        &        &        & \cmark & \cmark &        &        &        &        &        &        &        & \cmark &        &        &        & \cmark &        \\
\citet{DBLP:journals/corr/abs-2003-10699}  & Inf. Sci. & 2021 &        &        &        &        &        &        &        &        &        &        &        &        &        &        &        & \cmark &        &        &        &        &        &        & \cmark &        &        &        &        &        & \cmark &        \\
\citet{DBLP:journals/tkde/YuXCCHY24}       & TKDE     & 2022 &        &        &        &        &        &        &        &        &        & \cmark & \cmark &        &        &        &        & \cmark & \cmark &        &        &        &        &        &        & \cmark &        &        &        &        &        &        \\ \midrule
                                         &           &      & 17     & 14     & 10     & 9      & 9      & 8      & 7      & 7      & 4      & 4      & 3      & 3      & 3      & 3      & 3      & 40     & 25     & 15     & 13     & 9      & 4      & 4      & 7      & 21     & 13     & 5      & 4      & 3      & 10     & 1     \\
    \bottomrule
\end{tabular}
\end{table*}

%% file: tables/datasets_inpaper.tex
\begin{table}[h]
\centering
\caption{Dataset referencing in surveyed papers. The table reports the different ways datasets are cited, including references to the original data source, copies of the dataset, citations of the dataset’s original paper or other scholarly papers, missing references, and broken links.}
\begin{tabular}{l c}
\toprule
\textbf{Reference Type}                     & \textbf{\# Usages (Percentage)} \\ \midrule
Original data source     & 63 (35.2\%)                      \\
Copy a new version of the dataset         & 58 (32.4\%)                      \\
Dataset's original paper           & 27 (15.1\%)                      \\
Other scholarly papers             & 3 (1.7\%)                        \\
No reference                          & 19 (10.6\%)                      \\
Broken link             & 9 (5.0\%)                        \\ \midrule
\textbf{Total}                        & \textbf{179 (100\%)}             \\ \bottomrule
\end{tabular}
\label{tab:dataset_references}
\end{table}

%% file: sections/frameworks.tex
\subsection{Existing Approaches in Recommendation Frameworks}

Recommendation frameworks foster reproducibility and replicability of research, accelerate the development of recommender systems by eliminating the need for developers to reimplement algorithms from scratch, and contribute to more reliable and
impactful research. Many of these tools provide end-to-end architectures that manage the entire recommendation pipeline, while others focus on specific recommendation tasks.

\sloppy
Following the frameworks suggested by the ACM RecSys conference\footnote{\url{https://github.com/ACMRecSys/recsys-evaluation-frameworks}}, we analyze the most widely adopted ones. 
General-purpose end-to-end frameworks include \textbf{ClayRS}~\cite{DBLP:journals/is/LopsPMSS23}, \textbf{DaisyRec}~\cite{DBLP:conf/recsys/SunY00Q0G20}, \textbf{Elliot}~\cite{DBLP:conf/sigir/AnelliBFMMPDN21}, \textbf{LensKit}~\cite{DBLP:conf/cikm/Ekstrand20}, \textbf{RecBole}~\cite{DBLP:conf/cikm/ZhaoMHLCPLLWTMF21}, \textbf{Recommenders}~\cite{DBLP:conf/recsys/GrahamMW19}, and \textbf{RecPack}~\cite{DBLP:conf/recsys/MichielsVG22}, each designed to support different recommendation paradigms.
Other frameworks specialize in specific recommendation tasks. \textbf{Cornac}~\cite{DBLP:journals/jmlr/SalahTL20} targets multimodal recommendation, \textbf{FuxiCTR}~\cite{DBLP:conf/cikm/ZhuLYZH21} focuses on click-through rate prediction, and \textbf{ReChorus}~\cite{DBLP:conf/sigir/WangZMLM20} is tailored for sequential recommendation. In contrast, \textbf{RecList}~\cite{DBLP:conf/www/ChiaTBHK22} is not designed for pipeline management but serves as a behavioral testing framework.
One of the most recent frameworks, \textbf{Mab2Rec}~\cite{DBLP:conf/aaai/KadiogluK24}, addresses the lack of modularity in existing frameworks, which the authors argue hinders usability. To overcome this limitation, they introduce a modular framework specifically for multi-armed bandit recommenders.

Although these frameworks share models, algorithms, and design principles, they often function as isolated systems, each managing the entire recommendation pipeline. This results in monolithic architectures that limit interoperability and complicate the reproduction of data transformations across frameworks, posing a particular challenge for researchers aiming to experiment with models available only within specific ecosystems.
 
In addition, dealing with all elements of a recommendation pipeline requires constant maintenance and updating, leading to the potential oversight of some components that still need to be completed, updated, or even missing.
This limitation is further exacerbated by the lack of explicit interoperability between frameworks, which prevents the reuse of procedures already implemented in other projects. The resulting landscape is a collection of frameworks that re-implement existing solutions or, even worse, fail to implement some of the most common data handling strategies in recommendation. This is highlighted in~\Cref{tab:frameworks}.

\input{tables/frameworks}

Conversely, in \framework, we not only implemented the reported strategies, but thanks to our library’s modularity and the explicit interfaces for exporting datasets in a format compatible with other frameworks, these strategies can be easily integrated into existing frameworks, thus contributing to the overall reproducibility landscape of recommender systems.
It is essential to underline that these analyses could change as recommendation frameworks evolve over time.
RecList~\cite{DBLP:conf/www/ChiaTBHK22} was excluded from the analysis because it only handles the evaluation phase.





%% file: tables/frameworks.tex
\begin{table*}[]
\small
\caption{Comparison of \framework with existing recommendation frameworks based on data input/output, pre-filtering and dataset splitting strategies.}
\label{tab:frameworks}
\setlength{\tabcolsep}{3.2pt}
\rowcolors{18}{white}{gray!15}
\begin{tabular}{l@{\hskip 1em}lllrll@{\hskip 1em}llllllll@{\hskip 1em}lllllllllllllllllll}
\toprule
                                                                                  & \multicolumn{6}{c}{\textbf{Data Input/Output}}                                                                                                                                                              & \multicolumn{8}{c}{\textbf{Prefiltering}}                                                                                                                                                                                                                     & \multicolumn{19}{c}{\textbf{Splitting}}                                                                                                                                                                                                                                                                                                                                                                                                                                                                                                                                                                                                                           \\ \cmidrule(r){2-7}
                                                                                  \cmidrule(lr){8-15}
                                                                                  \cmidrule(l){16-34}
                                                                                  & \multicolumn{3}{c}{\multirow{2}{*}{\makecell{I/O   \\Formats}}}                   & \multicolumn{2}{c}{\multirow{2}{*}{\makecell{Avail. \\Data}}}                  & \multirow{3}{*}{\rotatebox{90}{Ext. frameworks output\hspace{0.95em}}} & \multicolumn{3}{c}{\multirow{2}{*}{\begin{tabular}[c]{@{}c@{}}Filter-\\ by-rating\end{tabular}}} & \multicolumn{5}{c}{\multirow{2}{*}{$k$-Core}}                                                                                                              & \multicolumn{6}{c}{\multirow{2}{*}{Temporal Hold-out}}                                                                                                                               & \multicolumn{12}{c}{Random}                                                                                                                                                                                                                                                                                                                                                                   & \multicolumn{1}{c}{\multirow{3}{*}{\rotatebox{90}{Pre-computed\hspace{4.9em}}}} \\
                                                                                  
                                                                                             \cmidrule(lr){22-33}
                                                                                  & \multicolumn{3}{c}{}                                                    & \multicolumn{2}{c}{}                                           &                                                         & \multicolumn{3}{c}{}                                                                             & \multicolumn{5}{c}{}                                                                                                                                       & \multicolumn{6}{c}{}                                                                                                                                                                 & \multicolumn{5}{c}{\textit{HO}}                                                                                                                              & \multicolumn{5}{c}{$K$\textit{-HO}}                                                                                                                          & \multicolumn{2}{c}{\textit{CV}}                                 & \multicolumn{1}{c}{}                                              \\
                                                                   \cmidrule(r){2-4} 
                                                       \cmidrule(l){5-6}                
                                                                    \cmidrule(lr){8-10}              \cmidrule(lr){11-15}
                                                                    \cmidrule(lr){16-21}
                                                       \cmidrule(lr){22-26}        \cmidrule(lr){27-31}      \cmidrule(lr){32-33}
                                                                                  & \rotatebox{90}{Tabular} & \rotatebox{90}{Inline} & \rotatebox{90}{JSON} & \rotatebox{90}{Built-in datasets} & \rotatebox{90}{Versioning} &                                                         & \rotatebox{90}{Numerical}    & \rotatebox{90}{Distributional}    & \rotatebox{90}{User Dist.}    & \rotatebox{90}{User} & \rotatebox{90}{Item} & \rotatebox{90}{Iterative} & \rotatebox{90}{Iter-$n$-rounds} & \multicolumn{1}{c}{\rotatebox{90}{Cold-Users}} & \rotatebox{90}{Fixed Timestamp} & \rotatebox{90}{By-Ratio Sys.} & \rotatebox{90}{By-Ratio User} & \rotatebox{90}{Leave-$1$-out} & \rotatebox{90}{Leave-$n$-in} & \rotatebox{90}{Leave-$n$-out} & \rotatebox{90}{By-Ratio Sys.} & \rotatebox{90}{By-Ratio User} & \rotatebox{90}{Leave-$1$-out} & \rotatebox{90}{Leave-$n$-in} & \rotatebox{90}{Leave-$n$-out} & \rotatebox{90}{By-Ratio Sys.} & \rotatebox{90}{By-Ratio User} & \rotatebox{90}{Leave-$1$-out} & \rotatebox{90}{Leave-$n$-in} & \rotatebox{90}{Leave-$n$-out} & \rotatebox{90}{$k$-folds Sys.} & \rotatebox{90}{$k$-folds User} & \multicolumn{1}{c}{}                                              \\ \midrule
ClayRS~\cite{DBLP:journals/is/LopsPMSS23}           &      \cmark               &         \cmark                &     \cmark                   &      1                &                                   &                            &         \cmark                                                &                              &                                   &                               &                      &                      &                           &                                 &                                                &                    &        \cmark                            &                  &                    &                    &                    &                \cmark                    &                  &                 &                  &                   &                               &                               &                               &                              &     \cmark                               &                                &                    \cmark                                                        \\

Cornac~\cite{DBLP:journals/jmlr/SalahTL20}                                        &               \cmark               &          \cmark                 &                      &                13                   &                            &                                                         & \cmark                   &                                   &                               & \cmark           & \cmark           &                           &                                 &                                                &                       &                               &                               &                               &                              &                               & \cmark                    & \cmark                    &                               &                              &                               &                               &                               &                               &                              &                               & \cmark                     &                                & \cmark                                                        \\

Elliot~\cite{DBLP:conf/sigir/AnelliBFMMPDN21}                                                                        &         \cmark                     &                        &                      &        1                           &                            &                                                         & \textbf{\cmark}          & \textbf{\cmark}               & \textbf{\cmark}           & \textbf{\cmark}  & \textbf{\cmark}  & \textbf{\cmark}       & \textbf{\cmark}             & \textbf{\cmark}                            & \textbf{\cmark}   & \textbf{}                     & \textbf{\cmark}           & \textbf{\cmark}           & \textbf{}                    & \textbf{\cmark}           & \textbf{}                     & \textbf{\cmark}           & \textbf{\cmark}           & \textbf{}                    & \textbf{\cmark}           & \textbf{\cmark}           &                               & \textbf{\cmark}           &                              & \textbf{\cmark}           & \textbf{}                      & \textbf{\cmark}            & \textbf{\cmark}                                               \\

FuxiCTR~\cite{DBLP:conf/cikm/ZhuLYZH21}                                &          \cmark                    &                        &                      &                  0                 &                            &                                                         &                              &                                   &                               &                      &                      &                           &                                 &                                                &                       &                               &                  &                &                   &                    &         \cmark                           &                    &                    &                  &                   &                               &                               &                               &                              &                               &                                &                     & \cmark                                                        \\


LensKit Python~\cite{DBLP:conf/cikm/Ekstrand20}                                &          \cmark               &                        &                      &                    1               &                            &                                                         &                              &                                   &                               &                      &                      &                           &                                 &                                                &                       &                               &                     &                     &                    &                     &                               &                     &                     &                    &                     &                               &                               &         \cmark                       &                               &          \cmark                      &            \cmark                     & \cmark                     & \cmark                                                        \\

Mab2Rec~\cite{DBLP:conf/aaai/KadiogluK24}                                &        \cmark                 &                        &                      &                   0                &                            &                                                         &                              &                                   &                               &                      &                      &                           &                                 &                                                &                       &                               &                 &                   &                    &                    &                               &                    &                     &                  &                 &                               &                               &                               &                              &                               &                                &                      & \cmark                                                        \\

Recommenders~\cite{DBLP:conf/recsys/GrahamMW19}                                &       \cmark                  &                        &                      &                 7                  &                            &                                                         &       \cmark                       &                                   &                               &        \cmark              &              \cmark        &              \cmark             &                                 &                                                &          \cmark             &             \cmark                  & \cmark                    &                     &                    &                  &           \cmark                    & \cmark                    &                    &                   &                     &                               &                               &                               &                              &                               &                                &                     & \cmark                                                        \\

RecBole~\cite{DBLP:conf/cikm/ZhaoMHLCPLLWTMF21}                                   &                         &                        &                      &                     66              &           \cmark                 &                                                         &             \cmark                 &                                   &                               &             \cmark         &     \cmark                 &             \cmark              &                    \cmark             &       \cmark                                         &                       &                               & \cmark                    & \cmark                    &                              & \cmark                    & \cmark                    & \cmark                    & \cmark                    &                              & \cmark                    &                               &                               &                               &                              &                               &                                &                                & \cmark                                                        \\

ReChorus~\cite{DBLP:conf/sigir/WangZMLM20}                                   &         \cmark                &                        &                      &                      3             &                            &                                                         &                              &                                   &                               &                      &                      &                           &                                 &                                                &                       &                               &                    &                    &                              &                    &                    &                   &                  &                              &                     &                               &                               &                               &                              &                               &                                &                                & \cmark                                                        \\


RecPack~\cite{DBLP:conf/recsys/MichielsVG22}                                   &        \cmark                 &                        &                      &                 11                  &                            &                                                         &                              &                                   &                               &           \cmark           &      \cmark                &                           &                                 &                                                &          \cmark             &                               & \cmark                    & \cmark                    &                              &                     & \cmark                    & \cmark                    &                     &                              &                    &                               &                               &                               &                              &                               &                                &                                & \cmark                                                        \\

\framework                                                                        &        \cmark                 &           \cmark             &        \cmark              &           18                        &                    \cmark        &               \cmark                                          & \textbf{\cmark}          &                &           & \textbf{\cmark}  & \textbf{\cmark}  & \textbf{\cmark}       & \textbf{\cmark}             & \textbf{\cmark}                            & \textbf{\cmark}   & \textbf{\cmark}                     & \textbf{\cmark}           & \textbf{\cmark}           & \textbf{\cmark}                    & \textbf{\cmark}           & \textbf{\cmark}                     & \textbf{\cmark}           & \textbf{\cmark}           & \textbf{\cmark}                    & \textbf{\cmark}           & \textbf{\cmark}           &    \cmark                           & \textbf{\cmark}           &        \cmark                      & \textbf{\cmark}           & \textbf{\cmark}                      & \textbf{\cmark}            & \textbf{\cmark}                                               \\ \bottomrule
\end{tabular}
\end{table*}

%% file: sections/xxx.tex
\section{\framework}

\begin{figure}
    \centering
    \includegraphics[width=\columnwidth]{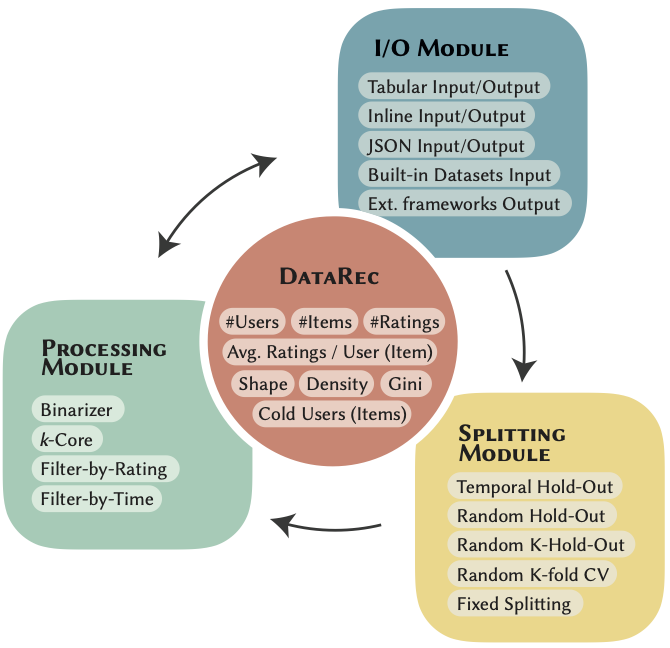}
    \caption{Overview of the \framework architecture. The \texttt{DataRec} class provides key dataset metrics and interacts with three modules: I/O for handling different data formats, processing for dataset transformations, and splitting for partitioning into training, validation, and test sets.}
    \label{fig:framework_overview}
    \vspace{-1em}
\end{figure}

\framework is an open-source Python library for handling recommendation datasets. In addition to providing the tools needed to manage datasets, \framework aims to ensure reproducibility, ease of use, data traceability, modularity, and interoperability. \Cref{fig:framework_overview} illustrates the fundamental components of the library that interact through the main class \texttt{DataRec}, which serves as the primary element for dataset manipulation.
In the following sections a more in-depth description of the \framework main functionalities: the input/output module, the \texttt{DataRec} Class, the processing and splitting modules, and the reproducibility strategies.

\subsection{The I/O Module}
This module handles the operations for reading and writing datasets in \framework. Datasets can be read from files in the most common formats and saved in the same formats. Moreover, to facilitate the export of datasets for direct use within other frameworks, export methods are provided for several frameworks.
To ensure traceability and versioning, \framework includes the most widely used datasets within its scope according to state-of-the-art practices identified in our analysis. For clarity, each of these aspects is discussed in detail below.

\subsubsection{Data Formats}
\framework supports the three main formats for representing recommendation datasets that arose from our analysis:
\begin{itemize}
    \item \textbf{Tabular:} Each row in the dataset represents a record of a user-item interaction, which may also include an explicit rating and a timestamp.
    \item \textbf{Inline:} This format represents implicit feedback between users and items. Each row represents a user's history, where the first element is the user and the following elements (in a variable number) represent the items in their history.
    \item \textbf{JSON:} This format facilitates the representation of unstructured content associated with interactions. It is structured as a sequence of objects, where each object represents a user-item interaction and its metadata, including an explicit rating and a timestamp.
\end{itemize}

For each of these formats, the library provides reading and writing methods, allowing conversion of a dataset from one format to another. During the input/output process, \framework uses the \texttt{RawData} class to represent the data. This class is a simplified version of the \texttt{DataRec} class and serves as a unified interface for I/O operations.

\adjustbox{center,margin=0.2cm,scale=0.9}{\begin{tcolorbox}[colback=black!10!white, colframe=black!75!white, title=Reading and Writing]
    \begin{verbatim}
from datarec.io import read_tabular, write_tabular

data = read_tabular(
path, sep='::', user_col=0, header=None)
                        
write_tabular(
data, path=output_path, sep='\t', header=None)
    \end{verbatim}
\end{tcolorbox}}

\subsubsection{Built-in Datasets}
Based on the findings of our survey, \framework incorporates the \textbf{16 most widely used recommendation datasets}~\cite{DBLP:conf/recsys/McAuleyL13, DBLP:conf/sigir/McAuleyTSH15, DBLP:conf/www/HeM16, hou2024bridging, DBLP:conf/emnlp/NiLM19, DBLP:conf/kdd/ChenHXGGSLPZZ19, DBLP:conf/kdd/ChoML11, DBLP:conf/recsys/2011hetrec, DBLP:conf/acl/WuQCWQLLXGWZ20, DBLP:journals/tiis/HarperK16, IJCAI16, Yelp}, each appearing in at least three different papers (along with other additional datasets). To mitigate versioning issues, we maintain the reference to the original data source whenever possible and implement available previous versions for backward compatibility. A summary of available datasets is available in~\Cref{tab:datasets-info}.

\input{tables/datasets_info}

When a dataset is directly available in \framework, the library allows the user to download it and use it in a ready-to-use \texttt{DataRec} object. 
For each version of each dataset available, \framework provides a public checksum to validate the correctness of the resource downloaded from the referenced source. 

\adjustbox{center,margin=0.2cm,scale=0.9}{\begin{tcolorbox}[colback=black!10!white, colframe=black!75!white, title=Built-in Dataset]
    \begin{verbatim}
from datarec.datasets import MovieLens

data = MovieLens(version='1m')
    \end{verbatim}
\end{tcolorbox}}

\subsubsection{Interoperability}
To facilitate the integration of library functionalities and results into existing recommendation frameworks, \framework provides methods for exporting a \texttt{DataRec} object in formats compatible with the following recommendation frameworks: 
\textbf{ClayRS}~\cite{DBLP:journals/is/LopsPMSS23},
\textbf{Cornac}~\cite{DBLP:journals/jmlr/SalahTL20},
\textbf{DaisyRec}~\cite{DBLP:conf/recsys/SunY00Q0G20},
\textbf{Elliot}~\cite{DBLP:conf/sigir/AnelliBFMMPDN21},
\textbf{LensKit}~\cite{DBLP:conf/cikm/Ekstrand20},
\textbf{RecBole}~\cite{DBLP:conf/cikm/ZhaoMHLCPLLWTMF21},
\textbf{ReChorus}~\cite{DBLP:conf/sigir/WangZMLM20},
\textbf{Recommenders}~\cite{DBLP:conf/recsys/GrahamMW19},
and \textbf{RecPack}~\cite{DBLP:conf/recsys/MichielsVG22}.
In addition, \framework includes utilities designed to streamline integration with each of these frameworks.

\adjustbox{center,margin=0.2cm,scale=0.9}{\begin{tcolorbox}[colback=black!10!white, colframe=black!75!white, title=Data Export]
    \begin{verbatim}
from datarec.io import FrameworkExporter
    
exporter = FrameworkExporter(output_path=path)
exporter.to_elliot(train, test, val)
    \end{verbatim}
\end{tcolorbox}}

\subsection{The \texttt{DataRec} Class}
The \texttt{DataRec} class is the main interface within the library for managing recommendation datasets. Internally, each dataset is stored as a pandas DataFrame object, which allows access to optimized methods for dataset management. The pandas DataFrame is structured as a table with at least two columns, one for the user and one for the item, with each record corresponding to a user-item interaction. If an interaction includes an explicit record, it is stored numerically in a third column. The fourth and final column, when provided, manages timestamps to capture temporal information.

Among the other functionalities of the \texttt{DataRec} class is the calculation of metrics for analyzing recommendation datasets, which have been shown to impact model performance~\cite{DBLP:journals/tmis/AdomaviciusZ12, DBLP:conf/sigir/DeldjooNSM20, DBLP:conf/wsdm/ChinCC22}. The implemented metrics include \textbf{space size}, \textbf{shape}, \textbf{density}, user and item \textbf{Gini coefficient}, and \textbf{average ratings} per user and per item.
Finally, users and items can be classified into four categories based on their \textbf{popularity} within the dataset: “most popular,” “popular,” “common,” and “long tail.” The classification is automatically determined by computing the quartile values of the user and item distributions.


\subsection{Processing Module}
This module implements the tools necessary for transforming a recommendation dataset. The implemented approaches are:

\begin{itemize}
    \item \textbf{Binarizer:} Converts explicit ratings into implicit feedback based on user-definable thresholds.
    \item \textbf{$k$-Core:} Filters out users and/or items with fewer than \( k \) recorded interactions. The iterative version (\( k \)-Core iterative) repeatedly applies this filtering until all remaining users and items meet the \( k \)-Core criterion. If convergence cannot be achieved, \framework allows setting a maximum number of iterations (\( \text{Iter-}n\text{-rounds} \)). Additionally, the framework includes an option to filter out cold users by retaining only those with a minimum level of prior interactions.
    \item \textbf{Filter-by-Rating:} This method eliminates user-item interactions when the preference score falls below a specified threshold. The threshold can be defined as: (1) a fixed numerical value (e.g., 2.5); (2) a global metric (e.g., the dataset-wide average rating); or (3) a user-specific measure (e.g., the user's mean rating).
    \item \textbf{Filter-by-Time:} Retains only interactions before or after a given time threshold. 
\end{itemize}


\input{figures/configuration}

\subsection{Splitting Module}

Once filtered, the data is partitioned using the splitting module, which supports several strategies for creating train-test sets:
\begin{itemize}
\item \textbf{Temporal Splitting}: Partitions interactions based on timestamps. This method supports fixed time thresholds, optimal cutoff points, or a hold-out (HO) mechanism. HO approaches can rely on specified ratios or removing the most recent items.
\item \textbf{Random Splitting}: Similar to temporal splitting, this method includes a hold-out option. Additionally, using $K$-repeated hold-out (K-HO) and cross-validation (CV), it can generate multiple train-test partitions.
\item \textbf{Pre-computed Splitting}: Uses predefined data partitions, which is helpful for benchmarks that require consistent train-test sets.
\end{itemize}
When applicable, these strategies are applied in a \textbf{user-stratified manner}, meaning that the splitting logic is executed on each user’s history individually rather than on the entire dataset. 


\subsection{Reproducibility}
To ensure reproducibility in not-deterministic procedures, \framework allows users to set a random seed. Additionally, \framework transparently tracks every operation performed on any \texttt{DataRec} object, maintaining a complete history. This history can be exported as a configuration file, which, when provided to \framework, enables all previous operations to be reproduced accurately. Furthermore, each operation records the checksum of the \texttt{DataRec} object’s state to verify its correctness. In~\Cref{listing:config}, an example configuration file is generated from a \texttt{DataRec} history.

%% file: tables/datasets_info.tex
\begin{table}[t]
    \centering
    \caption{Overview of the datasets implemented in \framework. The table lists the dataset name, file format, available versions, and source links.}
    \renewcommand{\arraystretch}{0.9}
        \begin{tabular}{lccl}
        \toprule
        \textbf{Datasets} & \textbf{Format} & \textbf{Versions} & \textbf{Source} \\
        \midrule
        Alibaba-iFashion~\cite{DBLP:conf/kdd/ChenHXGGSLPZZ19}& Inline & 2019  & \href{https://drive.google.com/drive/folders/1xFdx5xuNXHGsUVG2VIohFTXf9S7G5veq}{\textbf{[link]}}\tablefootnote{\url{https://drive.google.com/drive/folders/1xFdx5xuNXHGsUVG2VIohFTXf9S7G5veq}}  \\ \hline

        \multirow{2}{*}{Amazon Reviews~\cite{DBLP:conf/emnlp/NiLM19, DBLP:journals/corr/abs-2403-03952}}& Tabular & 2018 & \href{https://cseweb.ucsd.edu/~jmcauley/datasets/amazon_v2/}{\textbf{[link]}}\tablefootnote{\url{https://cseweb.ucsd.edu/~jmcauley/datasets/amazon_v2/}}  \\
         & Tabular & 2023 & \href{https://amazon-reviews-2023.github.io/}{\textbf{[link]}}\tablefootnote{\url{https://amazon-reviews-2023.github.io/}} \\ \hline

        CiaoDVD~\cite{DBLP:conf/asunam/GuoZTY14} & Inline & 2013 & \href{https://guoguibing.github.io/librec/datasets.html}{\textbf{[link]}}\tablefootnote{\url{https://guoguibing.github.io/librec/datasets.html}} \\ \hline

        Epinions~\cite{DBLP:conf/semweb/RichardsonAD03} & Inline & 2003 & \href{https://snap.stanford.edu/data/soc-Epinions1.html}{\textbf{[link]}}\tablefootnote{\url{https://snap.stanford.edu/data/soc-Epinions1.html}} \\ \hline

        Gowalla~\cite{DBLP:conf/kdd/ChoML11} & Inline & 2011 & \href{https://snap.stanford.edu/data/loc-gowalla.html}{\textbf{[link]}}\tablefootnote{\url{https://snap.stanford.edu/data/loc-gowalla.html}} \\ \hline

        Last.fm~\cite{DBLP:conf/recsys/2011hetrec} & Tabular & 2011 & \href{https://grouplens.org/datasets/hetrec-2011/}{\textbf{[link]}}\tablefootnote{\url{https://grouplens.org/datasets/hetrec-2011/}} \\ \hline

        Mind~\cite{DBLP:conf/acl/WuQCWQLLXGWZ20} & Tabular & 2020 & \href{https://msnews.github.io/}{\textbf{[link]}}\tablefootnote{\url{https://msnews.github.io/}} \\ \hline

        \multirow{2}{*}{MovieLens~\cite{DBLP:journals/tiis/HarperK16}}& Tabular & 2006 (1M) & \href{https://grouplens.org/datasets/movielens/1m/}{\textbf{[link]}}\tablefootnote{\url{https://grouplens.org/datasets/movielens/1m/}}  \\
         & Tabular & 2016 (20M) & \href{https://grouplens.org/datasets/movielens/20m/}{\textbf{[link]}}\tablefootnote{\url{https://grouplens.org/datasets/movielens/20m/}}   \\ \hline

        Tmall~\cite{IJCAI16} & Tabular & 2018 & \href{https://tianchi.aliyun.com/dataset/53?t=1716541860503}{\textbf{[link]}}\tablefootnote{\url{https://tianchi.aliyun.com/dataset/53?t=1716541860503}}  \\ \hline

        Yelp & JSON & 2023  & \href{https://www.yelp.com/dataset}{\textbf{[link]}}\tablefootnote{\url{https://www.yelp.com/dataset}}  \\       
        \bottomrule
        \end{tabular}
    \label{tab:datasets-info}
\end{table}

%% file: figures/configuration.tex
\begin{listing}[t]
\caption{\texttt{example\_configuration.yml}}
\vspace{-1em}
\begin{lstlisting}[language=yaml]

pipeline:
- name: load
  operation: MovieLens
  params:
    version: 1m
  checksum: c4d9eecfca2ab87c1945afe126590906
- name: process
  operation: Binarize
  params:
    threshold: 4
  checksum: 0c5a5e05efb79e561a2d9c6b087980ff
- name: process
  operation: UserItemIterativeKCore
  params:
    cores: 2
  checksum: ef1a1bca94111c164d17b03a1a5c1314
- name: split
  operation: RandomHoldOut
  params:
    test_ratio: 0.2
    val_ratio: 0.1
    seed: 42
  checksum:
    test: 81e4150e5230a15d7c0d97b3371ffab1
    val: 65c04aa6c326c832891dfe4815465855
    train: 9a6760e3da74a1984d6d0057739b14ad
- name: export
  operation: Elliot
  params:
    output_path: ./elliot/

\end{lstlisting}
\label{listing:config}
\end{listing}

%% file: sections/conclution.tex
\section{Conclusion and Future Work}
As observed from the related and recent literature on recommender systems, a critical limitation is represented by the lack of standardized procedures for recommendation data management, both in new proposed methodologies and in recommendation frameworks collecting them. To this end, we designed \framework to empower recommender systems developers and foster ongoing discussions and research aimed at enhancing dataset management. It offers a comprehensive suite of resources and tools for data management and processing, with an emphasis on ease of use, seamless integration, reproducibility, and traceability of data sources. Furthermore, to promote standardization and code reuse, \framework includes routines that facilitate integration with existing recommendation frameworks. As an open project, \framework is committed to continuous improvement, evolving to incorporate emerging datasets and novel strategies. In the near future, we plan to extend this work with more in-depth analyses and enhanced support for integrated side information management.